\documentclass[aps,prb,twocolumn,showpacs,groupedaddress,graphics]{revtex4}
\usepackage{graphicx}
\usepackage{epsfig}
\begin{document}
\bibliographystyle{apsrev}

\title{Shot Noise Suppression at Non-integer Conductance
Plateaus\\
in a Quantum Point Contact}
\author{N. Y. Kim}
\thanks{Corresponding author.}
\email[]{Email: nayoung@stanford.edu}
\author{W. D. Oliver}
\altaffiliation{Present address: MIT Lincoln Laboratory,
Lexington, Massachusetts, 02420}
\author{Y. Yamamoto}
\thanks{also at NTT Basic Research Laboratories,
3-1 Morinosato-Wakamiya Atsugi, Kanagawa, 243-01 Japan}
\affiliation{Quantum Entanglement Project, ICORP, JST,\\
E. L.Ginzton Laboratory, Stanford University, Stanford, California
94305}
\author{Y. Hirayama}
\affiliation{NTT Basic Research laboratories, 3-1
Morinosato-Wakamiya\\ Atsugi, Kanagawa, 243-01 Japan }
\date{November 19, 2003}

\begin{abstract}
\vspace{0.2in} We study non-equilibrium differential conductance
and current fluctuations in a single quantum point contact. The
two-terminal electrical transport properties -- differential
conductance and shot noise -- are measured at 1.5 K as a function
of the drain-source voltage and the Schottky split-gate voltage.
In differential conductance measurements, conductance plateaus
appear at integer multiples of $2e^2/h$ when the drain-source
voltage is small, and the plateaus evolve to a fractional of
$2e^2/h$ as the drain-source voltage increases. Our shot noise
measurements correspondingly show that the shot noise signal is
highly suppressed at both the integer and the non-integer
conductance plateaus. This main feature can be understood by the
induced electrostatic potential model within a single electron
picture. In addition, we observe the 0.7 structure in the
differential conductance and the suppressed shot noise around 0.7
 ($2e^2/h$); however, the previous single-electron model cannot explain the
0.7 structure and the noise suppression, suggesting that this
characteristic relates to the electron-electron interactions.
\end{abstract}

\maketitle

A quantum point contact (QPC) in a two-dimensional electron gas
(2DEG) system has been a prototypical device used to investigate
low-dimensional mesoscopic physics. The Landauer-B\"uttiker
formalism \cite{Buttiker92,Landauer92}, which interprets the
electrical transport in such devices, is the most widely used
theoretical model. By applying a negative voltage to
lithographically patterned Schottky gates on top of 2DEG,
additional spatial confinements can be achieved. Combinations of
QPCs form zero-dimensional quantum dots \cite{Kastner93} in which
single charge tunnelling occurs, and a single QPC defines
one-dimensional conducting channels \cite{Houten96}. In the latter
situation, the QPC becomes an electron waveguide that regulates
the number of transverse modes between electron reservoirs. As a
manifestation, a conductance trace consists of quantized steps in
integer multiples of the spin degenerate quantum unit of
conductance, $G_Q = 2 e^2 /h$, where $e$ is an electron charge and
$h$ is Planck's constant. Recently, the quantum modes of coherent
electrons under QPCs were imaged with atomic force microscopy
\cite{Topinka01}. An additional remarkable feature has been
identified around 0.7 $G_Q$, which is called the `` 0.7 structure
'' or `` 0.7 anomaly '' in the QPC conductance
\cite{Fitzgerald02}. Its physical origin is still under
investigation in terms of the interaction \cite{Thomas96} and spin
properties of electrons \cite{Cronenwett02} by means of
conductance.

The integer-plateau picture is true when a drain-source voltage
($V_{ds}$) is kept small. As $V_{ds}$ increases, the plateaus
evolve from integer units $n G_Q$ to non-integer units $(\beta +n)
 G_Q$, where $\beta$ is a fraction between 0 and 1 and $n$ is a
non-negative integer \cite{Patel91}. The transition of conductance
plateaus can be understood by a model of an electrostatic
potential which is a function of $V_{ds}$ \cite{Moreno92,
Kouwenhoven89}. Due to the discrepancy between the number of
allowed forward and backward transverse modes for a given finite
energy window, the location of quantized levels depends on the
degree ($\beta$) of the voltage drop across drain and source
sides.

Along with experimental and theoretical work on the conductance,
the current fluctuations have been studied as well with QPCs since
these fluctuations provides information that is not contained,
even in principle, in the conductance. Shot noise is the
non-equilibrium current fluctuation resulting from the stochastic
transport of quantized charge carriers. In mesoscopic conductors,
shot noise occurs due to the random partition of electrons by a
scatterer. Previous shot noise experiments with a QPC
\cite{Reznikov95,Kumar96} clearly showed that shot noise signals
agree well with a non-interacting theory, meaning that shot noise
is nearly zero at the integer conductance plateaus where electrons
are fully transmitted. We, however, have not yet found any
thorough noise studies on the characteristic around the 0.7
structure clearly seen in early noise data\cite{Liu98, Heiblum}
except as reported in the Ph.D. thesis \cite{Oliver02} regarding
to such structure.

In this Letter, we reexamine a single QPC and report our
experimental results on low frequency shot noise as well as
differential conductance at 1.5 K by sweeping both $V_{ds}$ and a
split-gate voltage $V_g$. We find a close connection between noise
and conductance data. Shot noise is suppressed when conductance
approaches quantized values of $G_Q$. Furthermore, the highly
reduced shot noise signals are resolved near other fractional
$G_Q$ regions and around 0.7 $G_Q$ for non-zero $V_{ds}$.

Our QPC devices were fabricated on a high mobility 2DEG formed in
an undoped GaAs/AlGaAs heterostructure. A back-gate field-effect
configuration allows us to tune the electron density in 2DEG
\cite{Hirayama02}. The average electron density is $~$ 2 $\times$
10$^{11}$ cm$^{-2}$. From the Hall bar pattern, the voltage drop
across the QPC can be probed so that QPC conductance was
experimentally extracted. Two external parameters --- $V_{ds}$ and
$V_g$ ---  were varied in both the differential conductance,
$g=dI/dV_{ds}$, and the low frequency two-terminal shot noise
measurements. A standard lock-in technique was used on the
differential conductance $g$ measurement. In order to improve the
signal-to-noise-ratio in the shot noise experiment, an ac
modulation lock-in technique and a resonant circuit were used
together with a home-built cryogenic low-noise preamplifier
\cite{Reznikov95, Liu98, Oliver99}. All measurements were
performed in a He$^3$ cryostat, whose base temperature was kept at
1.5 K.

\begin{figure}
\epsfig{figure=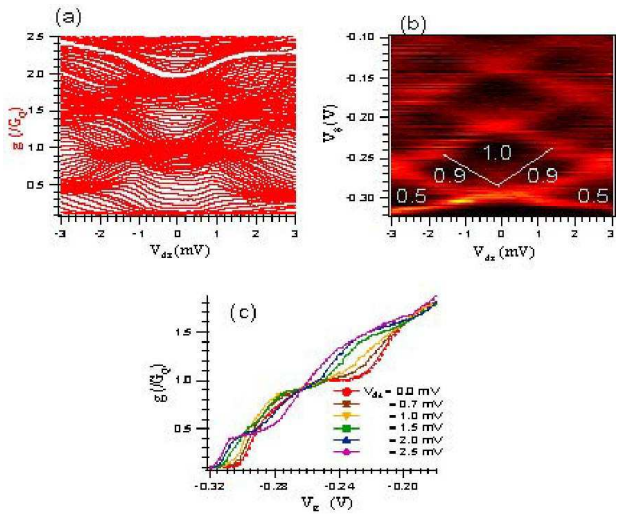, width=3.2in} \caption{(Color online) (a)
Differential conductance $g = dI/dV_{ds}$ as a function of
$V_{ds}$ normalized by $G_Q = 2e^2/h$ at 1.5 K. (b)
Transconductance $dg/dV_g$ obtained by mathematical analysis.
Black color indicates plateaus in (a) and red represents
transitions between plateaus. The first big black diamond region
can be divided into three sections; the top is from $G_Q$ plateau
and the bottom two are related to the 0.9 $G_Q$ plateaus. (c) The
traces of $g$ for various $V_{ds}$ versus $V_g$. \label{Fig:fig1}}
\end{figure}

The measured differential conductance $g$ with an ac bias voltage
$V_{ac} \sim $ 100 $\mu V$ is plotted as a function of $V_{ds}$
and $V_g$ in Fig.~\ref{Fig:fig1}(a). All data on each line are
taken at a different $V_g$, and all measured values are normalized
by $G_Q$. Dark regions are formed around the regions of plateaus.
Conductance flattens around $G_Q$ and 2 $G_Q$ along $V_{ds} \sim$
0, but away from $V_{ds} \sim$ 0, $g$ approaches plateaus at
different locations. Alternatively, Fig.~\ref{Fig:fig1}(c) clearly
illustrates that the first step appears below 0.5 $G_Q$ when
$V_{ds}$= - 2.5 mV. We compute transconductance $dg/dV_{g}$ by
differentiating $g$ in terms of $V_g$, and plot it in a
two-dimensional image graph (Fig. ~\ref{Fig:fig1}(b)). Here, black
areas correspond the plateaus due to the small difference between
traces along $V_g$ axis. In the first big diamond black area,
there is a V-shape red structure, which separates the 0.9 $G_Q$
structures from the $G_Q$ plateau.

Furthermore, we notice that the transition behavior is not
identical over the whole conductance values for finite $V_{ds}$.
Below $G_Q$, an additional shoulder structure around 0.7 $G_Q$ is
manifest and it moves to 0.9 $G_Q$, and then the plateau clearly
forms below 0.5 $G_Q$ at a large $V_{ds}$ . In contrast, above
$G_Q$, as $V_{ds}$ increases, no structure similar to the 0.7
anomaly is apparent and the plateau shows an increasing manner.
The appearance of the non-integer conductance plateaus in terms of
$V_{ds}$ is understood quantitatively by a $V_{ds}$-dependent
saddle-point potential model where the potential in a
two-dimensional $x$ and $y$ plane is given by\cite{Moreno92}
$$ U(x,y)  = U_0 (V_{ds}) + U_y y^2 - U_x x^2.$$
The first term in the right hand side contains the effect of a
non-zero $V_{ds}$ and it is written as:
$$U_0(V_{ds})= U_0 - \beta eV_{ds} + \gamma eV^2_{ds}/2,$$
where the coefficient $\beta$ is determined by the actual voltage
drop between the drain and source side, and $\gamma$ is related to
the trend of plateau movements as $V_{ds}$ gets
bigger\cite{Moreno92}.

\begin{figure}
\epsfig{figure=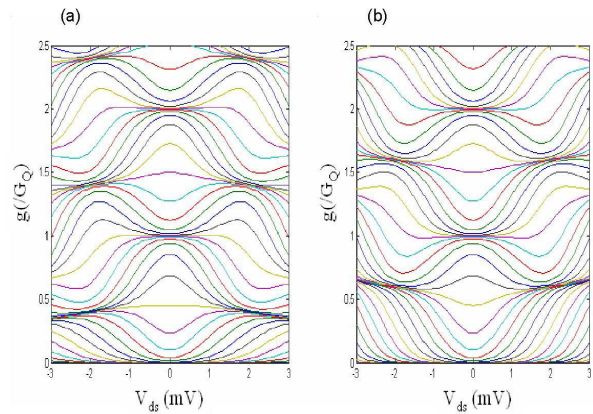,width = 3.2 in} \caption{The calculated
differential conductance $g$ based on the saddle-point potential
model which has both a linear and a quadratic term of $V_{ds}$.
The sign of $\gamma$ determines the manner of the plateau
evolution for a finite $V_{ds}$ : (a) a negative quadratic term
(decreasing plateaus)(b) a positive quadratic term (increasing
plateaus). Assuming a symmetric barrier, $\beta$ is chosen as 1/2
and $U_y/U_x$ is set to 2. Given a fixed value of $\gamma$, the
same plot as the measurement data cannot be generated.
\label{Fig:fig2}}
\end{figure}

Although the simulation result (Fig.~\ref{Fig:fig2}) show the
qualitative picture of plateau evolution in $V_{ds}$, it fails to
replicate the 0.7 $G_Q$ and 0.9 $G_Q$ structures, suggesting that
more complicated physical mechanism is involved under $G_Q$ and
especially around 0.7 $G_Q$ and 0.9 $G_Q$.

\begin{figure}
\epsfig{figure=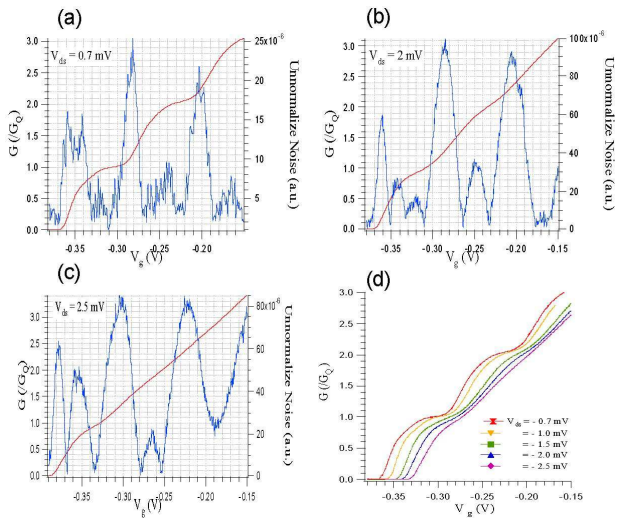,width=3.2in} \caption{(Color online) Shot
noise (blue: the right vertical axis scale) and conductance
$G=I/V_{ds}$ (Red: the left vertical axis scale) for various
$V_{ds}$ (a) 0.7 mV (b) 2 mV and (c) 2.5 mV. (d) $V_g$ dependence
of conductance $G$. For clarity, the traces are shifted along
$V_g$ axis. \label{Fig:fig3}}
\end{figure}

Following the differential conductance experiments, the low
frequency two-terminal shot noise measurements performed. In order
to extract the distinguishable shot noise signal from background
noise, $V_{ds}$ cannot be smaller than 500 $\mu V$.  Three
representative graphs are drawn as a function of $V_g$ in
Fig.~\ref{Fig:fig3}. Similar behaviors were observed in other
devices as well. No matter what value of $V_{ds}$ was applied, the
shot noise level was clearly minimal when conductance $G =
I/V_{ds}$ reached about $G_Q$ and 2 $G_Q$. The degree of the
suppression at 3 $G_Q$ became less smaller  for a large $V_{ds}$.
In the transient zones between the multiples of $G_Q$, the noise
characteristic was rather complex. Below the first plateau, the
noise suppression appeared around 0.6 $G_Q$ and 0.9 $G_Q$ until
$V_{ds} \sim$  1.5 mV (Fig.~\ref{Fig:fig3}(a)). As $V_{ds}$
further increased, these locations moved down to 0.5 $G_Q$ and 0.8
$G_Q$ (Fig.~\ref{Fig:fig3}(b)), and eventually the suppressed
noise was found only at 0.4 $G_Q$ for $V_{ds} >$ 2.5 mV
(Fig.~\ref{Fig:fig3}(c)). Unlikely, when $G$ is higher than $G_Q$,
only one additional noise reduction was found about 1.6 $G_Q$ or
1.7 $G_Q$ regardless of the magnitude of $V_{ds}$. Meanwhile, the
plateau structures in $G$ gradually washed out as $V_{ds}$
increased as shown in Fig.~\ref{Fig:fig3}(d).

\begin{figure}
\epsfig{figure=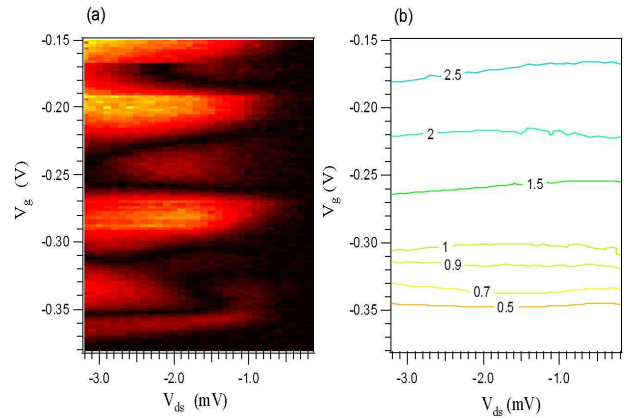,width=3.2in} \caption{(Color online) (a)
Shot noise is plotted as a function of both $V_{ds}$ and $V_g$.
Dark region represents suppressed noise level. (b) The contour map
of corresponding conductance $G$. The numbers represent normalized
conductance values by $G_Q$. \label{Fig:fig4}}
\end{figure}

Figure ~\ref{Fig:fig4} (a) exhibits the above observation of the
shot noise response as a function of $V_{ds}$ and $V_g$ in a
continuous manner. The black color depicts the base shot noise
level. Even though the occurrence of the suppressed shot noise can
be easily seen in units of $G_Q$, the actual plot contains other
noticeable features. The colored contour plot of conductance $G =
I/V_{ds}$(Fig.~\ref{Fig:fig4}(b)) helps us to see the relation of
$G$ and the shot noise. Again under $G_Q$, several black strips
are visible: The upper strip relates to the shot noise suppression
around $G_Q$, and the lower two ones start at the conductance
values 0.7 $G_Q$ and 0.9 $G_Q$. For a high $V_{ds}$, the shot
noise suppression occurs at less than 0.5 $G_Q$. The shot noise
signal in higher G has a rather simple pattern: The reduced noises
are observed around 1.6 or 1.7 $G_Q$ and 2 $G_Q$ as previously
stated.

We notice that the shot noise behavior in the transient zone
between the integer multiples of $G_Q$ shares some features with
the transconductance two-dimensional image plot
(Fig.~\ref{Fig:fig1}(b)). The peaks in the transconductance
correspond to the larger shot noise signals and the dark areas in
the transconductance match to the black strips in the shot noise
image. Moreover, both the transconductance and the shot noise
share common features for $G < G_Q$; 0.7 structure can be
distinctive and the location of the noise suppression and the new
plateaus in $dg/dV_g$ occur around 0.4 $G_Q$ as $V_{ds} >$ 2 mV.
Within the saddle-point potential model, $dg/dV_{g}$ is expressed
in terms of  $T_i(1 - T_i)$ where $T_i$ is the i-th
one-dimensional (1D) channel transmission probability. Since the
shot noise has a term of $T_i(1 - T_i)$ for a small energy window,
two quantities are closely related. It is not, however, obvious to
predict the response of the shot noise for a large $V_{ds}$
because the shot noise is obtained from the integral of the energy
dependent transmission probability. Qualitatively, the noise
suppression around the plateaus can be expected based on the fact
that the current fluctuations can be zero or low when the current
remains constant.

The different characteristics in both the transconductance and the
shot noise are observed in the region of $G < G_Q$ and $G
> G_Q$. This observation is certainly beyond the simple
saddle-point potential model in a single-particle approximation.
In particular, it is surprising to have the strongly suppressed
shot noise at 0.7 $G_Q$, meaning that electrons are regulated by a
certain governing physical mechanism. The possible factor relating
to the mechanism of the 0.7 anomaly would be the density of
electrons. The shot noise study in terms of the electron density
would provide more information to explore this question in the
future.

In conclusion, we have experimentally studied the low frequency
shot noise and the differential conductance with finite values of
$V_{ds}$. We showed that the main feature of shot noise
suppression in terms of $V_{ds}$ can be understood by the
differential conductance. However, the further investigation of
the properties of both the differential conductance and the shot
noise around the 0.7 structure should be needed in order to
establish better understandings.

We acknowledge the ARO-MURI grant DAAD19-99-1-0215 for supporting
this research.


\begin{thebibliography}{0}
\expandafter\ifx\csname natexlab\endcsname\relax\def\natexlab#1{#1}\fi
\expandafter\ifx\csname bibnamefont\endcsname\relax
  \def\bibnamefont#1{#1}\fi
\expandafter\ifx\csname bibfnamefont\endcsname\relax
  \def\bibfnamefont#1{#1}\fi
\expandafter\ifx\csname citenamefont\endcsname\relax
  \def\citenamefont#1{#1}\fi
\expandafter\ifx\csname url\endcsname\relax
  \def\url#1{\texttt{#1}}\fi
\expandafter\ifx\csname urlprefix\endcsname\relax\def\urlprefix{URL }\fi
\providecommand{\bibinfo}[2]{#2}
\providecommand{\eprint}[2][]{\url{#2}}

\end{thebibliography}


\begin{references}{}

\bibitem{Buttiker92} M. B\"uttiker, Phys. Rev. B{\bf 46}, 12485
(1992).

\bibitem{Landauer92} Th. Martin and R. Landauer, Phys. Rev. B {\bf
45} 1742 (1992).

\bibitem{Kastner93} M. A. Kastner, Physics Today {\bf 46}(1), 24
(1993).

\bibitem{Houten96} H. van Houten and C. Beenakker, Physics Today
{\bf 49}(7), 22 (1996).

\bibitem{Topinka01} M. A. Topinka, B. J. LeRoy, R. M. Westervelt,
S. E. J. Shaw, R. Fleischmann, E. J. Heller, K. D. Maranowski, and
A. C. Gossard, Nature {\bf 410}, 183 (2001).

\bibitem{Fitzgerald02} {\it Recent review} R. Fitzgerald,
Physics Today {\bf 55}(5), 21 (2002).

\bibitem{Thomas96} K. J. Thomas, J. T. Nicholls, M. Y. Simmons, M. Pepper,
D. R. Mace, and D. A. Ritchie, Phys. Rev. Lett. {\bf 77}, 135
(1996).

\bibitem{Cronenwett02} S. M. Cronenwett, H. J. Lynch, D. Goldhaber-Gordon,
L. P. Kouwenhoven, C. M. Marcus, K. Hirose, N. S. Wingreen, and V.
Umansky, Phys. Rev. Lett. {\bf 88}, 226805 (2002).

\bibitem{Patel91} N. K. Patel, J. T. Nicholls, L. Martin-Moreno, M. Pepper,
J. E. F. Frost, D. A. Ritchie, and G. A. C. Jones, Phys. Rev. B
{\bf 44}, 13549 (1991).

\bibitem{Moreno92} L. Martin-Moreno, J. T. Nicholls, N. K. Patel, and M. Pepper,
 J. Phys.:Condens. Matter {\bf 4}, 1323 (1992).

\bibitem{Kouwenhoven89} L. P. Kouwenhoven, B. J. van Wees, C. J.
P. M. Harmans, J. G. Williamson, H. van Houten, C. W. J.
Beenakker, C. T. Foxon, and J. J. Harris, Phys. Rev. B {\bf 39},
8040 (1989).

\bibitem{Reznikov95} M. Reznikov, M. Heiblum, H. Shtrikman, and D. Mahalu,
 Phys. Rev. Lett. {\bf 75}, 3340 (1995).

\bibitem{Kumar96} A. Kumar, L. Saminadayar, D. C. Glattli, Y. Jin, and
B. Etienne, Phys. Rev. Lett. {\bf 76}, 2778 (1996).

\bibitem{Liu98} R. C. Liu, B. Odom, Y. Yamamoto, and S. Tarucha,
 Nature {\bf 391}, 263 (1998).

\bibitem{Heiblum} M. Heiblum (Private communication)

\bibitem{Oliver02} W. D. Oliver, Ph.D. Dissertation (2002).

\bibitem{Oliver99} W. D. Oliver, J. Kim, R. C. Liu, and Y. Yamamoto,
 Science {\bf 284}, 299 (1999).

\bibitem{Hirayama02} Y. Hirayama and Y. Tokura (Private
communication)



%
%


\end{references}
\end{document}